\documentclass[aps,prb,reprint,twocolumn,superscriptaddress,floatfix,showpacs]{revtex4}

\usepackage[usenames,dvipsnames]{pstricks}

\usepackage{epsfig,graphics,graphicx}

\usepackage{amsmath,amsfonts,amssymb,latexsym}

\usepackage{dsfont}

\begin{document}

\title{Quantum and classical correlations and Werner states in finite spin linear arrays}

{\frenchspacing

\author{P.~R.~Wells~Jr.} \email{wells@if.ufrj.br}

\affiliation{Instituto de F{\'\i}sica, Universidade Federal do Rio de Janeiro Cx.P. 68528, 21945-970, Rio de Janeiro, RJ, Brazil}

\author{C.~M.~Chaves}

\affiliation{Centro Brasileiro de Pesquisas F{\'\i}sicas, Rua Xavier Sigaud 150, 22290-180, Rio de Janeiro, RJ, Brazil}

\author{J.~A.~e.~Castro}

\affiliation{Instituto de F{\'\i}sica, Universidade Federal do Rio de Janeiro Cx.P. 68528, 21945-970, Rio de Janeiro, RJ, Brazil}

\author{Belita~Koiller}

\affiliation{Instituto de F{\'\i}sica, Universidade Federal do Rio de Janeiro Cx.P. 68528, 21945-970, Rio de Janeiro, RJ, Brazil}

\date{\today}

\begin{abstract}

Pairwise quantum correlations in the ground state of a N-spins  antiferromagnetic chain described by the Heisenberg model with nearest neighbor exchange coupling are investigated. By varying a single coupling between two neighboring sites it is possible to drive spins from entangled to disentangled states, reversibly.
For even N the two-spin density matrix is written in the form of a Werner state, allowing identification of the weight parameter with the usual spin-spin correlation function $\langle S_i^z \, S_j^z \rangle = \Gamma_{ij}$. The  correlation functions show universal behavior in the $\Gamma$-dependence. 
This study presents a concrete possibility for the practical demonstration of entanglement control, opening alternatives for probing non-classical correlations and the realization of Werner states in familiar condensed matter systems. All required fabrication and measurement ingredients are currently available.

\end{abstract}


\maketitle

}

\section{Introduction}

Quantum behavior at the microscopic scale is well established both theoretically and experimentally. Strictly quantum phenomena observed so far have confirmed the validity of quantum mechanics postulates, in particular its non-local character.\cite{aspect1981} Entanglement is certainly one of the most intriguing  and basic of such quantum phenomena, as it is a fundamental resource in performing most tasks in quantum information and quantum computation.\cite{nielsen}  Experiments demonstrating and quantifying entanglement have been realized in the frameworks of photons \cite{walborn2006, almeida2007environment, jimenez2009determining} and atoms, \cite{anderson1995observation, yu2009sudden}
and preparation of specified entangled states has also been studied in model systems.\cite{PhysRevA.64.012313}

Separable quantum states, expected to show strictly classical behavior, can exhibit quantum correlations other than entanglement. A possible measure of quantum correlations is the quantum discord. \cite{PhysRevLett.88.017901, jphys-a34-6899-hend-vedral, PhysRevB.78.224413, PhysRevA.77.042303, sarandy:022108, PhysRevA.81.042105, ferraro2010almost, PhysRevLett.105.190502, PhysRevA.82.032112, PhysRevLett.105.095702, soares2010, PhysRevLett.107.070501, PhysRevLett.107.140403} The behavior of entanglement and quantum discord under variable external magnetic fields and nonzero temperatures, as well as of the spin-spin coupling  anisotropy have been discussed in the literature.\cite{PhysRevLett.87.017901, PhysRevLett.88.107901, PhysRevA.81.044101, RevModPhys.84.1655}

We have investigated related questions in short magnetic chains of spin-1/2 particles within the isotropic Heisenberg model. An important aspect of such chains is the tunability \cite{footnote1} of the coupling ($J$) between spins in neighboring sites.\cite{Petta30092005, PhysRevLett.107.030506, Nowack02092011} From the analysis of pairwise correlations in the ground state (GS) of even-numbered chains, we have identified entanglement-disentanglement (E-D) transitions as the coupling $J$ in one bond is varied.

The two-spin states under consideration are closely related to exact realizations of the so-called Werner state (WS).\cite{werner1989quantum, PhysRevA.66.062315, PhysRevA.70.022321, jin2010quantum} A WS is a one-parameter state that represents a superposition of a mixed component with a pure one, and is useful as it simulates the effect of the environment in destroying coherence and entanglement in a quantum system. \cite{PhysRevA.84.042316, PhysRevA.86.032110}

The WS parameter $p$ gives the relative weight of the two components, and drives the E-D transition. For the systems under consideration, we have found that $p = - 4 \Gamma_{ij}$, where $\Gamma_{ij} = \langle S_i^z S_j^z \rangle$ is the familiar spin-spin correlation function. In addition, we have obtained universal behavior for quantum and classical correlations as functions of $\Gamma_{ij}$.

Despite some difficulties, the experimental preparation of a Werner state is a subject of considerable interest. The first proposal in this respect involved photons in the context of quantum optics.~\cite{PhysRevA.66.062315} Other possibilities involve atoms in cavity electrodynamics~\cite{jin2010quantum} and nuclear magnetic resonance systems.\cite{PhysRevA.85.022329} We propose here the implementation in solid state systems, which would represent simpler preparation (the system is kept in the ground state) schemes associated with measurement methods already available or to be soon available. The case of $N=4$ is analyzed in detail and it is argued that that it presents practical advantages.

This paper is organized as follows. In Sec.~\ref{sec:theory} we present our model system and briefly review the measures quantifying correlations and entanglement used in our work. Sec.~\ref{sec:results} focuses on the $N=4$ case. Sec.~\ref{sec:werner} discusses the realization of WS and the identification of the control parameter $p$ with the spin-spin correlation function. Our final conclusions and experimental perspectives are given in Sec.~\ref{sec:concl}.

\section{\label{sec:theory} Model system and theoretical background}

We start by describing our model system and briefly reviewing the theoretical background used in our work.
\begin{equation}
H = \sum_{\langle i,j \rangle}^{N} J_{ij} \, \mathbf{S}_i \cdot \mathbf{S}_j \, ,
\label{eq:heisenberg}
\end{equation}
where the summation is over first neighbors of an open chain with $N$ spins, and the exchange couplings are positive ($J_{ij} > 0$).
The symmetries of the Hamiltonian (\ref{eq:heisenberg}) imply that the two-spin density matrix for any pair of spins, obtained by tracing over the $N - 2$ remaining ones, in the $S_z$ basis $\{ | \! \! \uparrow \uparrow \rangle, | \! \! \uparrow \downarrow \rangle, | \! \! \downarrow \uparrow \rangle, | \! \! \downarrow \downarrow \rangle \}$, has the general form \cite{sarandy:022108, PhysRevA.69.022311}
\begin{equation}
\rho_{ij} = \left(
\begin{array}{cccc}
 a & 0 & 0 & 0 \\
 0 & b_1 & z & 0 \\
 0 &  z& b_2 & 0\\
 0 & 0 & 0 & d
\end{array} \right) \, \, ,
\label{eq:densitymatrix}
\end{equation}
with $\text{Tr} \rho_{ij} = 1$, $b_1 b_2 \ge |z|^2$ and $a d \ge 0$. \cite{PhysRevA.81.042105}
Regarding correlation measures, several quantities have been proposed to describe correlations in quantum systems; we summarize below those adopted in our study.

Two interacting spins $i$ and $j$ share information that is quantified by the quantum mutual information (QMI),
\begin{equation}
{\cal{I}} (\rho_{ij}) = {\cal{S}}(\rho_i) + {\cal{S}}(\rho_j) - {\cal{S}}(\rho_{ij}) \, \, ,
\label{eq:qmi-v1}
\end{equation}
which includes all (quantum and classical) correlations between spins $i$ and $j$, and where ${\cal{S}} (\rho) = - \, \text{Tr} \, \rho \log_{2} \rho$ is the Von Neumann entropy.
We define Quantum correlation $(Q^C)$ as
\begin{equation}
Q^C (\rho_{ij}) = \min_{ \{ \Pi_{m}^{j} \} } \delta(\rho_{ij}) \, \, ,
\label{eq:qcorr}
\end{equation}
where the minimization is with respect to all possible measurement basis sets (projectors) $\{ \Pi_{m}^{j} \}$. $\delta(\rho_{ij})$ is the Quantum Discord, \cite{PhysRevLett.88.017901} defined as $\delta(\rho_{ij}) = {\cal{I}}(\rho_{ij}) - {\cal{J}}(\rho_{ij})$, where ${\cal{J}}(\rho_{ij}) = {\cal{S}}(\rho_i) - {\cal{S}}(\rho_{i|j})$, and $S (\rho_{i|j})$ is the quantum conditional entropy, i.e., the entropy of spin $i$ when the state of spin $j$ is known. This minimization amounts to finding the measurement basis that minimally disturbs the system, extracting information about spin $j$ with minimum disturbance on spin $i$. Defining classical correlation $(C^C)$ as \cite{jphys-a34-6899-hend-vedral, PhysRevA.77.042303, sarandy:022108}
\begin{equation}
C^C (\rho_{ij}) = \max_{ \{ \Pi_{m}^{j} \}} {\cal{J}}(\rho_{ij}) \, \, ,
\label{eq:ccorr}
\end{equation}
it is possible to decompose the  total correlation as a sum of quantum and classical contributions (since we use projective measurements):
\begin{equation}
{\cal{I}}(\rho_{ij}) = Q^C (\rho_{ij}) + C^C (\rho_{ij}) \, \, .
\end{equation}
The individual components $Q^C$ and $C^C$ are not symmetric in $i\leftrightarrow j$, since ${\cal{J}}$ is not, while their sum ${\cal{I}}$ is symmetric.
Finally, the usual measurement for the degree of entanglement between two spins is the concurrence,  \cite{PhysRevLett.78.5022, PhysRevLett.80.2245}
\begin{equation}
{\cal{C}} = \max \{ 0, \Lambda \} \, \, ,
\label{eq:conc}
\end{equation}
where $\Lambda = \sqrt{\lambda_1} - \sqrt{\lambda_2} - \sqrt{\lambda_3} - \sqrt{\lambda_4}$, and the four numbers $\lambda_{i = 1, \cdots ~ 4}$ are the eigenvalues in decreasing order of the operator $\rho_{ij}\tilde{\rho}_{ij}$, with $\tilde{\rho}_{ij} = ( \sigma_y \otimes \sigma_y ) \rho_{ij} ( \sigma_y \otimes \sigma_y )$, and $\sigma_y$ is the second Pauli matrix.
The concurrence for systems described by the density matrix of Eq.~(\ref{eq:densitymatrix}) can be directly obtained from Eq.~(\ref{eq:conc}) and is given by
\begin{equation}
{\cal{C}} = 2 \max ( 0, |z| - \sqrt{ad} ) \, \, .
\label{eq:conc-analytic}
\end{equation}
We refer below to all quantities, ${\cal C}_{ij} \equiv {\cal C} ( \rho_{ij} )$, $I_{ij} \equiv I ( \rho_{ij} )$, $C^C_{ij} \equiv C^C ( \rho_{ij} )$ and $Q^C_{ij} \equiv Q^C ( \rho_{ij} )$, as \textit{correlations}, whenever there is no ambiguity.

\section{\label{sec:results} Four-spins chain}

%-------figure-----interacting spins chain in the first scheme here------------

\begin{figure}[b!]
\includegraphics[width=.5\columnwidth]{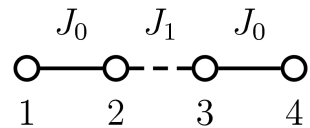}
\caption{\label{fig:3sc} Spin configuration with four spins, with a variable coupling
between spins $2$ and $3$. The other couplings are fixed to $J_0$.}
\end{figure}

%--------------------------------------

%

\begin{figure}[]

\includegraphics[width=\columnwidth]{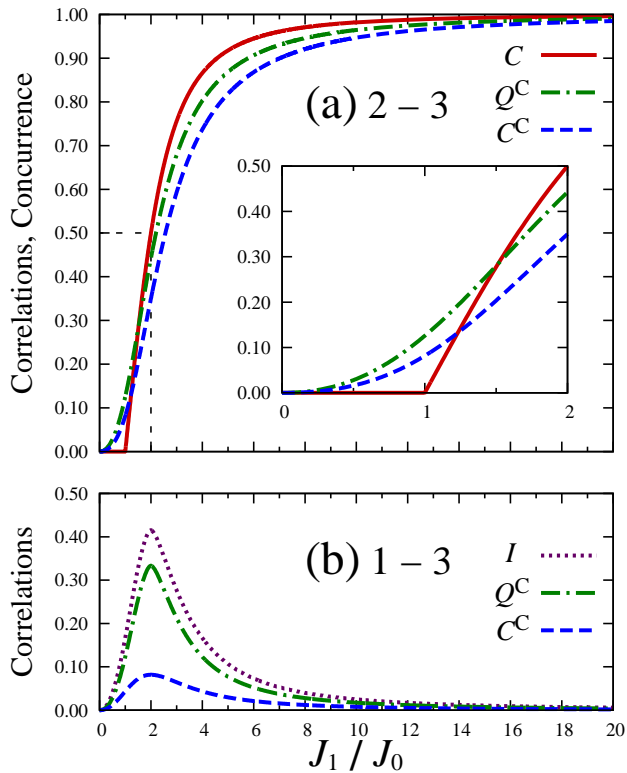}

\caption{\label{fig:4spins} (Color online) Correlations as a function of $\xi  = J_1/J_0$ for selected spin pairs of the GS of the four-spins chain depicted in Fig.~\ref{fig:3sc}. (a) The full, dash-dotted and dashed lines are respectively for ${\cal{C}}$, $Q^C$ and $C^C$ of pair $2-3$. The inset shows the region $0 \leq \xi \leq 2$. (b) Dotted, dash-dotted and dashed lines refer to total ($I$), quantum ($Q^C$) and classical ($C^C$) correlations for the pair $1-3$, whose concurrence is zero in the full range. See text for details.}

\end{figure}

The one-dimension Heisenberg model (\ref{eq:heisenberg}) has been studied in a wide variety of contexts; in particular, the qualitatively distinct behavior of even and odd-$N$ spin AF chains is already well established. \cite{{PhysRev.135.A640, PhysRevLett.90.047901, PhysRevB.68.134417, chaves:104410, chaves:032505}}

For AF coupling, the lowest state is in the lowest $S_{z}^{T} = \sum_i^N S_i^z$ subspace, so even-$N$ chains have  a non-degenerate $S_{z}^{T} = 0$ ground state (GS). Otherwise, odd-$N$ chains GS are doubly-degenerate, and we do not consider this possibility here.

We consider the $N = 4$ chain schematically shown in (Fig.~\ref{fig:3sc}); all couplings are AF, and the central exchange $J_1$ is varied while the others are kept fixed and equal to $J_0$. We are interested in the behavior of correlations as a function of the ratio $\xi = J_1 / J_0$. By explicitly writing the full GS density matrix on the $S_{z}^{T} = 0$ four-spins basis, we obtain the following symmetry relations for the reduced density matrices: $\rho_{12} = \rho_{34}$, $\rho_{13} = \rho_{24}$ and $\rho_{14} = \rho_{23}$ for all $\xi$. Also, $\rho_{12} (\xi) = \rho_{23} (4 / \xi)$.

Fig.~\ref{fig:4spins} shows correlations as functions of $\xi$ for pairs (a) $2-3$ and (b) $1-3$: ${\cal C}$, $Q^C$ and $C^C$ present similar behavior, and for each pair $C^C_{ij}(\xi)<Q^C_{ij}(\xi)$ for all $\xi$. The concurrence, however, shows an E-D transition: ${\cal C}_{23} = 0$ for $0 \leq \xi < 1$, but for $\xi > 1$, ${\cal C}_{23}$ increases rapidly, reaches the value $1/2$ at $\xi = 2$, and asymptotically approach unity. This limit corresponds to a singlet formed by spins $2$ and $3$. Independently and simultaneously, spins $1$ and $4$ also tend to form a singlet, as anticipated by the non-intuitive symmetry $\rho_{23} (\xi) = \rho_{14} (\xi)$. The singlet formed by spins 1 and 4 is a manifestation of \textit{long distance entanglement},\cite{PhysRevLett.96.247206, PhysRevB.82.140403} obtained here for correlations other than entanglement.

The correspondence $\xi \leftrightarrow 4 / \xi$ between pairs $1-2$ and $2-3$ allows inferring the behavior of ${\cal C}_{12}$ from that of ${\cal C}_{23}$: ${\cal C}_{12}$ is unit at $\xi = 0$, and decreases as $\xi$ increases, with ${\cal C}_{12}=0$ for $\xi>4$, in a manifestation of the so called monogamous behavior:~\cite{PhysRevA.61.052306,PhysRevA.69.022309} correlations in pairs $1-2$ and $3-4$ which are maximal for $J_1 = 0$, are gradually lost when $J_1$ increases, and correlations are built over pairs $2-3$ and $1-4$.

As regards to the other correlations, $Q^{C}_{12} \, ( = Q^{C}_{34})$ is unity for $\xi = 0$, and decreases gradually, approaching to zero asymptotically for large $\xi$, and $Q^{C}_{23} \, ( = Q^{C}_{14})$ increases from zero at $\xi = 0$ and asymptotically tends to one. The same type of behavior is obtained for $C^{C}_{12} \, (= C^{C}_{34})$ and $C^{C}_{23} (= C^{C}_{14})$ [See Fig.~\ref{fig:4spins}(a)]. We have $C^{C} \leq Q^{C}$, but ${\cal{C}}$ can be less or greater than $C^{C}$ and $Q^{C}$. Fig.~\ref{fig:4spins}(b) shows that quantum and classical correlations are nonzero for the pairs $1-3$ and $2-4$, with a maximum at $\xi = 2$, while ${\cal{C}}_{13} = {\cal{C}}_{24} = 0$ over the full range.

\section{\label{sec:werner} Werner states}

As noted in Sec.~\ref{sec:results}, an E-D transition may occur for specific pairs of the chain at finite values of $\xi$, e.g, ${\cal C}_{23}$ in Fig.~\ref{fig:4spins}(a) vanishes for $0<\xi<1$.

This behavior is characteristic of a WS,~\cite{PhysRevLett.88.017901}
\begin{equation}
\rho_W = \frac{1 - p}{4} \, \mathds{1}_4 + p \, | \Psi_{-} \rangle \langle \Psi_{-} | \, \, ,
\label{eq:werner-def}
\end{equation}
where  $\mathds{1}_4$ is the identity matrix in the four-dimensional space and $| \Psi_{-} \rangle = (| 0 1 \rangle - | 1 0 \rangle )/\sqrt{2}$ is a Bell state. The unspecified parameter $p$ drives an E-D  transition at $p = 1/3$.  

Its relation to the GS of an even-$N$ AF chain, for which $S_z^T = 0$, emerges from rotational and time reversal symmetries of the spin Hamiltonian, which imply that the elements of $\rho_{ij}$ in Eq.~(\ref{eq:densitymatrix}) may be written as \cite{PhysRevA.69.022311}
\begin{equation}
a = d = \frac{1}{4} + \Gamma_{ij} \, , \, b_1 = b_2 = \frac{1}{4} - \Gamma_{ij} \, , \, z = 2 \Gamma_{ij} \, \, .
\label{eq:coefs-n-par}
\end{equation}
Here,
\begin{equation}
\Gamma_{ij} = \langle S_i^z S_j^z \rangle \, \,
\label{eq:corrfunc}
\end{equation}
is the spin-spin correlation function ($-1/4 \le \Gamma_{ij} \le 1/4$ for spins 1/2). The results above are obtained regardless the size of the chain, the only restriction is $N$ even; so they remain valid for more general situations than the example of Sec.~\ref{sec:results}. For al pairs $i,j$ Eq.~(\ref{eq:densitymatrix}) gives
\begin{equation}
\rho^{N \rm even} = \frac{ 1 + 4 \Gamma }{4} \, \mathds{1}_4 - 4 \Gamma \, | S \rangle \langle S | \, \, ,
\label{eq:werner-def-2}
\end{equation}
where, without ambiguity, the $i,j$ labels are omitted. Here, $| S \rangle = \left( \vert \! \! \uparrow \downarrow \rangle - \vert \! \! \downarrow \uparrow \rangle \right)/\sqrt{2}$ represents a singlet, a maximally entangled two-spin state.
It then follows  that when  $ \Gamma > -1/12$ the concurrence vanishes and the state is separable, as for $p<1/3$. Eqs. (9) and (12) allow identifying $p = - 4 \Gamma$, attributing physical significance to $p$ in AF chains.~\cite{footnote2} Given that $\Gamma$ is a function of $\xi$, it is possible to reversibly control E-D transitions in this system by tuning $J_1$.

\subsection{\label{sub34} Universal behavior for quantum and classical correlations}

\begin{figure}[]
\includegraphics[width=.95\columnwidth]{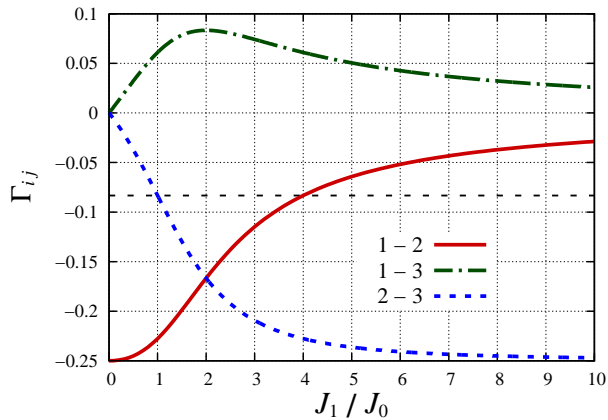}
\caption{\label{fig:gammaxi} (Color online) Spin-spin correlation function, $\Gamma_{ij}$, as a function of the coupling ratio for the four-spins chain in Fig~\ref{fig:3sc}. We find $\Gamma_{13} > 0$, ${\cal{C}}_{13} = 0$, in agreement with Fig.~\ref{fig:4spins}(b). The horizontal dashed line, given by $\Gamma = -1/12$, crosses $\Gamma_{12}$ and $\Gamma_{23}$, respectively, at the transition points $\xi^{12}_{C} = 4$ and $\xi^{23}_{C} = 1$.}
\end{figure}

The spin-spin correlation functions for all spin pairs in Fig~\ref{fig:3sc} are given in Fig.~\ref{fig:gammaxi} as a function of the exchange ratio parameter.  All pairs, $1-2$, $2-3$, $3-4$ and $1-4$, for which $\Gamma$ is negative, undergo an E-D transition.
The effective coupling between second-neighbors spins is ferromagnetic, leading to positive $\Gamma$ and null entanglement for pairs $1-3$ and $2-4$.
\begin{figure}[]
\includegraphics[width=\columnwidth]{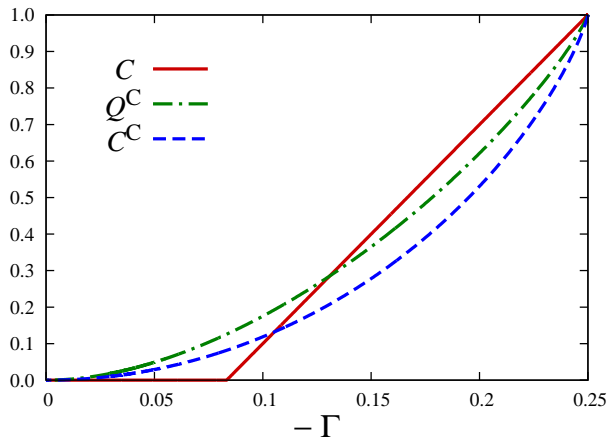}
\caption{\label{fig:corrxgamma} (Color online)  Correlations and concurrence as a function of minus the spin-spin correlation function in the $\Gamma < 0$ region for a four-spins chain. The curves are universal for pairs $1-2$, $2-3$, $3-4$, and $1-4$.}
\end{figure}
When the relation $\Gamma = f(\xi)$ is invertible for a pair $i-j$, as is the case for $\Gamma_{12}$ and $\Gamma_{23}$ for $N=4$ (See Fig.~\ref{fig:gammaxi}), correlations for  $i-j$ may be explicitly obtained as functions of $\Gamma$. This means that, for pairs with $\Gamma < 0$, the correlations dependence on $\Gamma$ is universal. For example, $C_{12}$ and $C_{23}$ [Fig.~\ref{fig:4spins}(a)] collapse into
\begin{equation}
{\cal{C}} = 6 \max \{ 0, -\Gamma - 1/12 \} \, \, , \, \, \Gamma < 0 \, \, ,
\label{eq:concws}
\end{equation}
given by the full line of Fig.~\ref{fig:corrxgamma}. The quantum and classical correlations for these pairs,  represented in Fig.~\ref{fig:corrxgamma}, are also universal: $Q^{C}_{12}$ and $Q^{C}_{23}$ [Fig.~\ref{fig:4spins}(a)] collapse into the dash-dotted line of Fig.~\ref{fig:corrxgamma}, while $C^{C}_{12}$ and $C^{C}_{23}$ into the dashed line.

\section{\label{sec:concl} Discussion and Conclusions}

The WS realization suggested here shows attractive features with respect to actual realizations or other proposals in the literature. \cite{PhysRevA.66.062315,jin2010quantum, PhysRevA.70.022321, PhysRevLett.109.066403} The reason for that is the fact that it involves the ground state of a well studied system in Condensed Matter Physics. This means that the superposition state of Eq.~(\ref{eq:werner-def}) does not require preparation other than cooling down the system to the ground state. This allows driving E-D transitions reversibly by varying the coupling $J_1$. Another important result in our work is the identification of the control parameter $p$ with a familiar quantity in the theory of magnetism. At this point, it is appropriated to comment on the feasibility of the proposed realization of the WS in short magnetic chains. 
Monoatomic chains of magnetic atoms have already been fabricated for $N =$ 3 to $10$ atoms, \cite{hirjibehedin2006spin} and individual atomic magnetizations $\langle S^z_i \rangle$ have been recently measured via spin-resolved scanning tunneling spectroscopy. \cite{nphys2299-Khajetoorians}

An important recent development is the demonstration by Loth et \textit{al} \cite{Loth13012012} that $N\sim 6 \to 12$ antiferromagnetic single Fe atoms chains may reasonably act as classical bits. Given that the electronic magnetic moment of transition metals is $S\gtrsim 2 \gg 1/2$, a semi-classical treatment is adequate for Fe, providing classical bits. The use of spin 1/2 electrons in similar chains, where we have shown that entanglement control is possible, may be useful in defining reliable spin-based qubits.~\cite{PhysRevLett.90.047901}
The measurement of $\langle S_i^z S_j^z \rangle$ requires simultaneous measurements in both spins $i$ and $j$. Such measurements have been performed recently in a double quantum dot.\cite{Nowack02092011}
Candidate physical systems for experimental verification of WS behavior in spin chains include: (i) donor-bound electron spins in a donors array precisely positioned in Si. \cite{kane1998} Such arrays with up to four P donors are within fabrication capabilities, as discussed in Ref.~\onlinecite{doi:10.1021/nl2025079}; (ii) Quantum dots arrays with single electron occupation each, \cite{PhysRevA.57.120, PhysRevB.59.2070} as four and five-dots arrays fabrication was recently reported.\cite{PhysRevLett.107.030506, Awschalom08032013} 
A recently considered related model, probing spin-spin entanglement transitions, refers to two Kondo spin chains, each one formed by an impurity and a bulk, coupled together through a Ruderman-Kittel-Kasuya-Yosida (RKKY) interaction between the impurities.~\cite{PhysRevLett.109.066403}
Although our results hold in principle for all even values of $N$, we argue here that the $N=4$ chain presents a particular symmetry, namely that all correlations in the internal pair $2-3$ with varying coupling are equal to the ones for the external pair $1-4$. This more distant pair is likely to present instrumentation advantages in terms of performing spin-spin correlations measurement: the coupling $J_1$ between internal spins in systems (i) and (ii) above could in principle be controlled through nano-electrodes conveniently aligned with respect to the target bond, while the external spins could be more easily accessed by a measurement apparatus.

In summary, we have studied spin chains in the context of classical and quantum pair correlations. We identify spin pairs that undergo reversible E-D transitions. The two-spin density matrix of these pairs is of Werner type and we find the connection of the parameter $p$ with the familiar spin-spin correlation function $\Gamma$. Universal behavior is obtained for quantum correlations as functions of $\Gamma$. Recent experiments on related systems point to the perspective that observation of E-D transitions (and WS) in simple and well studied condensed matter system seems to be accessible with current, or soon to be available, nano-processing capabilities.

\begin{acknowledgments}

We acknowledge Thereza Paiva, Andr\'e Saraiva, Ruynet de Matos Filho and Juan Carlos Retamal for valuable suggestions and interesting conversations. This work was partially supported by the Brazilian agencies CNPq, CAPES and FAPERJ, under the programs PAPDRJ, Cientista do Nosso Estado and Pensa-Rio.

\end{acknowledgments}


\begin{thebibliography}{10}
\bibitem{footnote1}%
  \BibitemOpen
  \bibinfo {note} {By varying gate voltages that raise or lower the
  electrostatic barrier between the electrons in the pair.}%
  \bibAnnoteFile{Stop}{footnote1}%
\end{thebibliography}
\end{document}